# A solar C/O and sub-solar metallicity in a hot Jupiter atmosphere

Michael R Line[1,2], Matteo Brogi[3,4,5], Jacob L. Bean[6], Siddharth Gandhi[3,4], Joseph Zalesky[1], Vivien Parmentier[7], Peter Smith[1], Gregory N. Mace[8], Megan Mansfield[9], Eliza M.-R. Kempton[10], Jonathan J. Fortney[11], Evgenya Shkolnik[1,2], Jennifer Patience[1], Emily Rauscher[12], Jean-Michel Desert´ [13], & Joost P. Wardenier[7]

[1]School of Earth and Space Exploration, Arizona State University, Tempe, AZ 85287, USA; mrline@asu.edu

[2]NASA Astrobiology Institute, Virtual Planetary Laboratory Team, Seattle, Washington, USA

[3]Department of Physics, University of Warwick, Coventry CV4 7AL, UK

[4]Centre for Exoplanets and Habitability, University of Warwick, Coventry, UK [5]INAF–Osservatorio Astrofisico di Torino, Turin, Italy

[6]Department of Astronomy and Astrophysics, University of Chicago, 5640 South Ellis Avenue, Chicago, IL 60637, USA

[7]Atmospheric, Oceanic, and Planetary Physics, Clarendon Laboratory, Department of Physics, University of Oxford, Oxford OX1 3PU, UK

[8]Department of Astronomy, University of Texas at Austin, 2515 Speedway, Austin, TX, USA

[9]Department of Geophysical Sciences, University of Chicago, 5734 S. Ellis Ave., Chicago, IL 60637

[10]Department of Astronomy, University of Maryland, 4296 Stadium Drive, College Park, MD 20742, USA

[11]Department of Astronomy and Astrophysics, University of California, Santa Cruz, CA 95064,
USA

[12]Department of Astronomy, University of Michigan, 1085 South University Avenue, Ann Arbor, MI 48109, USA

[13]Anton Pannekoek Institute for Astronomy, University of Amsterdam, Science Park 904, 1098 XH Amsterdam, The Netherlands

Measurements of the elemental abundances in exoplanet atmospheres, specifically of carbon (C) and oxygen (O), provide insight into the planet formation processes and evolution[1,2]. Assessing the C and O inventory in the hottest exoplanets ("hot Jupiters") requires bounded abundance determinations on the dominant molecular reservoirs, water ($H_2O$) and carbon-monoxide (CO). Previous observations of hot Jupiters have been able to provide bounded constraints on either $H_2O$ (from the Hubble Space Telescope)[3,4,5] or CO[6,7], but not both. Here we report observations of the hot Jupiter, WASP-77Ab, which enable atmospheric gas volume mixing ratio constraints on both the $H_2O$ and CO ($9.5\times10^{-5}$-$1.5\times10^{-4}$ and $1.2\times10^{-4}$-$2.6\times10^{-4}$, respectively). From these bounded constraints, we are able to derive the atmospheric C/H ($0.35^{+0.17}_{-0.10}\times$Solar) and O/H ($0.32^{+0.12}_{-0.08}\times$Solar) abundances and the corresponding atmospheric carbon-to-oxygen ratio (C/O=$0.59\pm0.08$). The sub-solar C and O abundances are suggestive of a metal-depleted atmosphere relative to expectations based on extrapolation from the solar system planets. The C/O constraint rules out planet formation scenarios that result in C-rich planetary atmospheres. Within the context of past inferences, these results point to a diversity of planetary atmospheric compositions and formation processes.

We observed the day side hemisphere of the tidally-locked transiting hot Jupiter WASP-77Ab (1740 K, 1.21 $R_J$, 1.76 $M_J$, 1.36 day period[8]) for a single 4.7 hour continuous time-series sequence on 14 December, 2020 with the Immersion GRating INfrared Spectrometer (IGRINS[9]) at the Gemini-South (GS) Observatory located on Cerro Pachon, Chile. Owing to the broad wavelength range (1.43 - 2.42 $\mu$m over 54 spectral orders), high spectral resolution (R~45,000), and sensitivity (SNR~180-270/resolution-element), IGRINS on GS is particularly sensitive to the molecular lines from multiple carbon, nitrogen, oxygen, and sulfur bearing species (see Methods). Seventy-nine separate spectra (140 s/spectrum) were obtained during the pre-eclipse portion (Fig. 1a) of the orbit when the hottest planetary hemisphere is present, covering a phase ($\varphi$) range between 0.32 and 0.47 (where $\varphi$=0 is transit and 0.5 is occultation/secondary eclipse). The IGRINS Pipeline Package[10] is used for the basic data reduction, spectral extraction, and initial wavelength calibrations, with additional reduction steps described in the Methods. When observing from the ground, it is necessary to remove the contaminating effects of Earth's atmosphere. Leveraging the rapidly changing Doppler shift of the planetary lines (~140 km $s^{-1}$ over the observing window) compared to the relatively stationary telluric (0 km $s^{-1}$) and stellar lines (~0.2 km $s^{-1}$), a principle component analysis (PCA, see methods) can be used to identify and remove the dominant time dependent contaminating sources[11], leaving the planetary signal largely unscathed.

Removal of the telluric contamination also removes any continuum level information in the planet-to-star flux ratio[7]. In order to extract meaningful information from data processed in this way[12,13], we must first cross-correlate (CC) the data with model templates. Using a set of representative thermal emission models that include the dominant absorbers expected at these temperatures and over the IGRINS wavelength range (primarily, $H_2O$ and CO), we cross-correlate as a function of velocity against each processed spectrum. Provided the model is an adequate template, the CC function (CCF) for each spectrum reaches its maximum at the planetary velocity (a sum of the system velocity and orbital velocity) at that specific orbital phase, and hence, trace out a CC trail in velocity[14]. The CC trail is clearly visible in Fig. 1b, corresponding to the appropriate planetary velocity components, demonstrating that we are detecting the planetary atmosphere as the planet orbits the star. We further leverage the circular orbital geometry, which predicts the phase-dependent line-of-sight Doppler shift given the planetary orbital ($K_p$) and relative system velocities ($\Delta V_{sys}$), to determine the total atmospheric signal detection by summing over the CCF at each phase[12,13] (see Methods). Fig. 1c shows the total atmospheric thermal emission crosscorrelation signal-to-noise, peaking at an S/N=12.8 very near (offset by ~-7 km s$^{-1}$ $\Delta V_{sys}$, see Methods) the anticipated pair of velocities, clearly indicating a strong detection of atmospheric thermal emission.

The next step is to identify the specific trace molecular species (the bulk atmosphere is predominately $H_2$/He) present in the spectrum and retrieve their absolute abundances. To do this we perform an automated procedure via a Bayesian inference (retrieval) scheme[7] (see Methods). This approach simultaneously optimizes the volume mixing ratios for each trace species ($\log_{10}(n_i)$, $i$=$H_2O$, CO, $CH_4$, $H_2S$, $NH_3$, and HCN as well as a CO isotopic abundance ratio-see Methods), the vertical temperature structure, the planetary orbital and system velocities, and nuisance parameters to account for uncertainties in the reported transit timing and possible signal stretching due the PCA analysis (see Methods). This method accounts for all of the degeneracies that arise amongst the the multiple overlapping molecular lines and absorption strength with atmospheric temperature gradient and permits for absolute molecular abundance determinations[5,7].

We achieve bounded constraints for $\log_{10}(n_{H2O})$ and $\log_{10}(n_{CO})$ (Fig. 2a) and only upper limits for the other species (see Methods), consistent with atmospheric chemical composition predictions under typical (~solar elemental composition, thermochemical equilibrium) assumptions[15] (Fig. 1c). We are also able to retrieve a bounded constraint on the CO isotopic abundance ratio (see Methods for a discussion). These data prefer a monotonically decreasing relatively cool temperature profile (Fig. 2d), suggestive of an atmosphere that either has a fairly efficient day-to-night atmospheric circulation (the predicted day-side temperature for poor circulation planets should follow the hotter inverted predicted profile in Fig. 2d) or lacks high altitude UV/optical absorbers (e.g., metal hydrides and oxides), possibly indicative of night-side condensation (cold-trapping) of refractory species[16].

The intrinsic elemental abundances in a planetary atmosphere are illuminating quantities because they are diagnostic of both atmospheric chemical processes and formation conditions. Furthermore, C and O account for ~70%[17] of the total "metals" (e.g., any species heavier than H, He) in a typical solar-like composition gas, and are hence, good tracers for the metal enrichment of an atmosphere. Since $H_2O$ and CO are the dominant C and O bearing molecules in this atmosphere (with relatively low abundance upper limits on the other major C and O bearing molecules-see Methods), are expected to be largely unperturbed by disequilibrium chemistry mechanisms at these temperatures[15], and are expected to be homogeneous with altitude over the pressures probed by typical observations[15] (Fig. 2c), we can convert them directly into the elemental oxygen ($n_O = n_{CO} + n_{H_2O}$) and carbon ($n_C = n_{CO}$) abundances. It is customary to normalize the elemental abundances relative to hydrogen ($n_i/n_H$), relative to that in the sun ($[X/H]:=\log_{10}((n_X/n_H)/(n_X/n_H)_{sun})$)[18] to facilitate comparisons with other astrophysical bodies in a common abundance reference frame. We find the elemental abundances in the atmosphere of WASP-77Ab to be $[C/H]=-0.46^{+0.17}_{-0.16}$, $[O/H]=-0.49^{+0.14}_{-0.12}$, $[(C+O)/H]=-0.48^{+0.15}_{-0.13}$, and a ratio of carbon to oxygen, $C/O=0.59\pm0.08$ (the Solar value is 0.55) (Fig. 2b) (all error bars reflect the 68% confidence interval). We also retrieve a subterrestrial $^{12}C/^{13}C$ abundance ratio (10.2-42.6 at 68% confidence, terrestrial value is 89), but see methods for a discussion and interpretation of the CO isotopic abundance constraint. Fig. 3a summarizes the [C/H] and [O/H] compared to the solar system giant planets. **With these abundance measurements we find that the C and O abundances are both sub-solar/stellar (WASP-77A has been measured to have a solar [Fe/H],[8]) by 2.7 and 3.5$\sigma$, respectively, and fall below the solar system values**, suggestive of different conditions for WASP-77Ab's atmosphere formation than for our own solar system giants.

When and where a planet forms within the protoplanetary disk, the relative role of solid versus gas accretion, and chemical processing ultimately dictate the observed atmospheric compositions, resulting in numerous potential outcomes for the elemental enrichment and abundance ratios. From the plethora of planet formation models, a few broad predictions have emerged for Jovian planet ($M>0.3M_J$) atmosphere compositions[2]: (i) Formation beyond the major ice lines ($H_2O$, $CO$, $CO_2$) and subsequent inwards migration after disk dissipation leads to elevated ($>0.8$) C/O and relatively low metal enrichment[18,19,20], (ii) Formation and migration *within* a disk results in substantial oxygen rich planetesimal pollution, resulting in low ($<0.5$) C/O, and elevated metal enrichment, with [X/H] decreasing with increasing planet mass[1,19], and (iii) Pebble accretion and drift[20,21] can result in both high C/O and super-solar metallicities.

It is with the sheer numbers of exoplanets that we can quantitatively test specific formationto-atmosphere hypotheses. By combining our abundance measurements with the solar system C abundances and $H_2O$-based O abundances from low resolution Hubble Space Telescope (HST) observations[4] we can glean some insight into the diversity of planet formation outcomes (Fig. 3b). The Solar system carbon abundances (black diamonds) follow a decreasing trend (dotted line) with increasing planet mass[22]. The low resolution HST-based oxygen abundances[4] show virtually no trend with mass but span ~0.03-300×Solar enrichments, though the constraints are rather coarse **(typically ~order-of-magnitude abundance precisions[4], compared to the factor of 1.5 obtained in this work)** for most objects. The WASP-77Ab C and O abundances both fall below the solar/stellar composition line and below the trend line predicted by the solar system, along with a few other O-based hot Jupiter abundances. These relatively low overall enrichments and ~solar C/O are not consistent with the above broad predictions–e.g., ~solar C/O but low metal enrichment. Instead, a possible formation scenario consistent with the measured abundances could be that the planetary core accreted its atmosphere interior to the major ice lines with O rich but C depleted gas (possibly due to sequestration into refractory grains), a relative lack of planetesimal bombardment,which would deliver both C and O, post atmosphere accretion, and little to no dissolution of the core metals into the atmosphere (Fig. 3b).

The challenge in connecting giant planet atmosphere compositions to their formation conditions is formidable. Over the past decades the planetary science community has made substantial progress on this front, starting with carbon and nitrogen abundances in the solar system planets, to order-of-magnitude oxygen abundance constraints in hot Jupiter atmospheres, a stringent upper limit on the Jovian oxygen abundance from JUNO[23], to now the first precision carbon *and* oxygen abundance measurements in exoplanets, advancing theory with each new measurement paradigm. Improvement in our understanding of how atmospheres came to be and how they evolve will continue as we push towards higher precision abundance measurements of more targets and for more elements from both ground **(dozens of planets are accessible with the level of precision presented here with current instruments)**- and space-based platforms **(e.g., The James Webb Space Telescope)**, ultimately paving the way for understanding our own Solar system's formation history in the galactic context.

**Acknowledgements** M.R.L, J.J.F, J.L.B, and P.S. acknowledge support from NASA XRP grant 80NSSC19K0293. M.R.L. and E.S. acknowledge support from the Nexus for Exoplanet System Science and NASA Astrobiology Institute Virtual Planetary Laboratory (No. 80NSSC18K0829). M.B. and S.G. acknowledge support from the UK Science and Technology Facilities Council (STFC) research grant ST/S000631/1. J.Z. acknowledges support from the NASA FINESST grant 80NSSC19K1420. E.M-R.K. &. E.R. thank the Heising-Simons Foundation for support. J.P.W. acknowledges support from the Wolfson Harrison UK Research Council Physics Scholarship and the UK Science and Technology Facilities Council (STFC). This work used the Immersion Grating Infrared Spectrometer (IGRINS) that was developed under a collaboration between the University of Texas at Austin and the Korea Astronomy and Space Science Institute (KASI) with the financial support of the Mt. Cuba Astronomical Foundation, of the US National Science Foundation under grants AST-1229522 and AST-1702267, of the McDonald Observatory of the University of Texas at Austin, of the Korean GMT Project of KASI, and Gemini Observatory

**Author Contributions** M.R.L conceived of the idea , performed the data analysis and modeling, and wrote the manuscript. J.Z. (PI) and M.R.L. wrote the original IGRINS proposal. M.B. provided guidance on the cross-correlation analysis and conceptual framework. J.L.B. provided guidance on the context of the results. S.G. performed an independent Bayesian analysis to confirm the result. G.N.M. ran the PLP pipeline and also assisted in the IGRINS specific observational setup. V.P., P.S., G.M., M.M., E.M.-R.K., J.J.F., E.S., J.P., E.R., J-M.D, J.P.W., and L.P. helped with the original proposal/and or provided valuable insight/comments on the manuscript or through discussions.

**Competing Interests** The authors declare that they have no competing financial interests.

**Correspondence** Correspondence and requests for materials should be addressed to M.R.L (email: mrline@asu.edu).

# References

1. Mordasini, C., van Boekel, R., Molliere, P., Henning, T. & Benneke, B. The Imprint of Exoplanet Formation History on Observable Present-day Spectra of Hot Jupiters. *Astrophys. J.* **832**, 41 (2016).

2. Madhusudhan, N. Exoplanetary Atmospheres: Key Insights, Challenges, and Prospects. *Ann. Rev. Astron. Astrophys.* **57**, 617-663 (2019).

3. Tsiaras, A. et al. A Population Study of Gaseous Exoplanets. *Astron. J.* **155**, 156 (2018).

4. Welbanks, L. et al. Mass-Metallicity Trends in Transiting Exoplanets from Atmospheric Abundances of $H_2O$, Na, and K. *Astrophys. J. Lett.* **887**, L20 (2019).

5. Gandhi, S., Madhusudhan, N., Hawker, G. & Piette, A. HyDRA-H: Simultaneous Hybrid Retrieval of Exoplanetary Emission Spectra. *Astron. J.* **158**, 228 (2019).

6. Pelletier, S. *et al.* Where is the Water? Jupiter-like C/H ratio but strong $H_2O$ depletion found on $\tau$ Bootis b using SPIRou.¨ *arXiv e-prints* arXiv:2105.10513 (2021).

7. Brogi, M. & Line, M. R. Retrieving Temperatures and Abundances of Exoplanet Atmospheres with High-resolution Cross-correlation Spectroscopy. *Astron. J.* **157**, 114 (2019).

8. Maxted, P. F. L. *et al.* WASP-77 Ab: A Transiting Hot Jupiter Planet in a Wide Binary System. *Pub. Astron. Soc. Pac..* **125**, 48 (2013).

9. Park, C. *et al.* Design and early performance of IGRINS (Immersion Grating Infrared Spectrometer). In Ramsay, S. K., McLean, I. S. & Takami, H. (eds.) *Ground-based and Airborne Instrumentation for Astronomy V*, vol. 9147 of *Society of Photo-Optical Instrumentation Engineers (SPIE) Conference Series*, 91471D (2014).

10. Mace, G. *et al.* IGRINS at the Discovery Channel Telescope and Gemini South. In Evans, C. J., Simard, L. & Takami, H. (eds.) *Ground-based and Airborne Instrumentation for Astronomy VII*, vol. 10702 of *Society of Photo-Optical Instrumentation Engineers (SPIE) Conference Series*, 107020Q (2018).

11. de Kok, R. J. *et al.* Detection of carbon monoxide in the high-resolution day-side spectrum of the exoplanet HD 189733b. *Astron. J.* **554**, A82 (2013).

12. Brogi, M. *et al.* The signature of orbital motion from the dayside of the planet $\tau$ Bootis b. *Nature* **486**, 502-504 (2012).

13. Birkby, J. L. *et al.* Detection of water absorption in the day side atmosphere of HD 189733 b using ground-based high-resolution spectroscopy at 3.2 $\mu$m. *Mon. Not. R. Astron. Soc.* **436**, L35-L39 (2013).

14. Snellen, I. A. G., de Kok, R. J., de Mooij, E. J. W. & Albrecht, S. The orbital motion, absolute mass and high-altitude winds of exoplanet HD209458b. *Nature* **465**, 1049-1051 (2010).

15. Moses, J. I. Chemical kinetics on extrasolar planets. *Philosophical Transactions of the Royal Society of London Series A* 372, 20130073-20130073 (2014).

16. Parmentier, V., Fortney, J. J., Showman, A. P., Morley, C. & Marley, M. S. Transitions in the Cloud Composition of Hot Jupiters. *Astrophys. J.* **828**, 22 (2016).

17. Asplund, M., Grevesse, N., Sauval, A. J. & Scott, P. The Chemical Composition of the Sun. *Ann. Rev. Astron. Astrophys.* **47**, 481-522 (2009).

18. Öberg, K. I., Murray-Clay, R. & Bergin, E. A. The Effects of Snowlines on C/O in Planetary Atmospheres. *Astrophys. J. Lett.* **743**, L16 (2011).

19. Madhusudhan, N., Amin, M. A. & Kennedy, G. M. Towards Chemical Constraints on HotJupiter Migration. *Astrophys. J. Lett.* **794**, L12 (2014)

20. Madhusudhan, N., Bitsch, B., Johansen, A. & Eriksson, L. Atmospheric signatures of giantexoplanet formation by pebble accretion. *Mon. Not. R. Astron. Soc.* **469**, 4102-4115 (2017).

21. Booth, R. A., Clarke, C. J., Madhusudhan, N. & Ilee, J. D. Chemical enrichment of giantplanets and discs due to pebble drift. *Mon. Not. R. Astron. Soc.* **469**, 3994-4011 (2017).

22. Kreidberg, L. *et al.* Clouds in the atmosphere of the super-Earth exoplanet GJ1214b. *Nature* **505**, 69-72 (2014).

23. Li, C. *et al.* The water abundance in Jupiter's equatorial zone. *Nat. Astron.* **4**, 609-616 (2020).

24. Burrows, A. & Sharp, C. M. Chemical Equilibrium Abundances in Brown Dwarf and Extrasolar Giant Planet Atmospheres. *Astrophys. J.* **512**, 843-863 (1999).

25. Atreya, S. K. *et al.* The Origin and Evolution of Saturn, with Exoplanet Perspective. *arXiv e-prints* arXiv:1606.04510 (2016).

26. Thorngren, D. & Fortney, J. J. Connecting Giant Planet Atmosphere and Interior Modeling: Constraints on Atmospheric Metal Enrichment. *Astrophys. J. Lett.* **874**, L31 (2019).

Main Figures

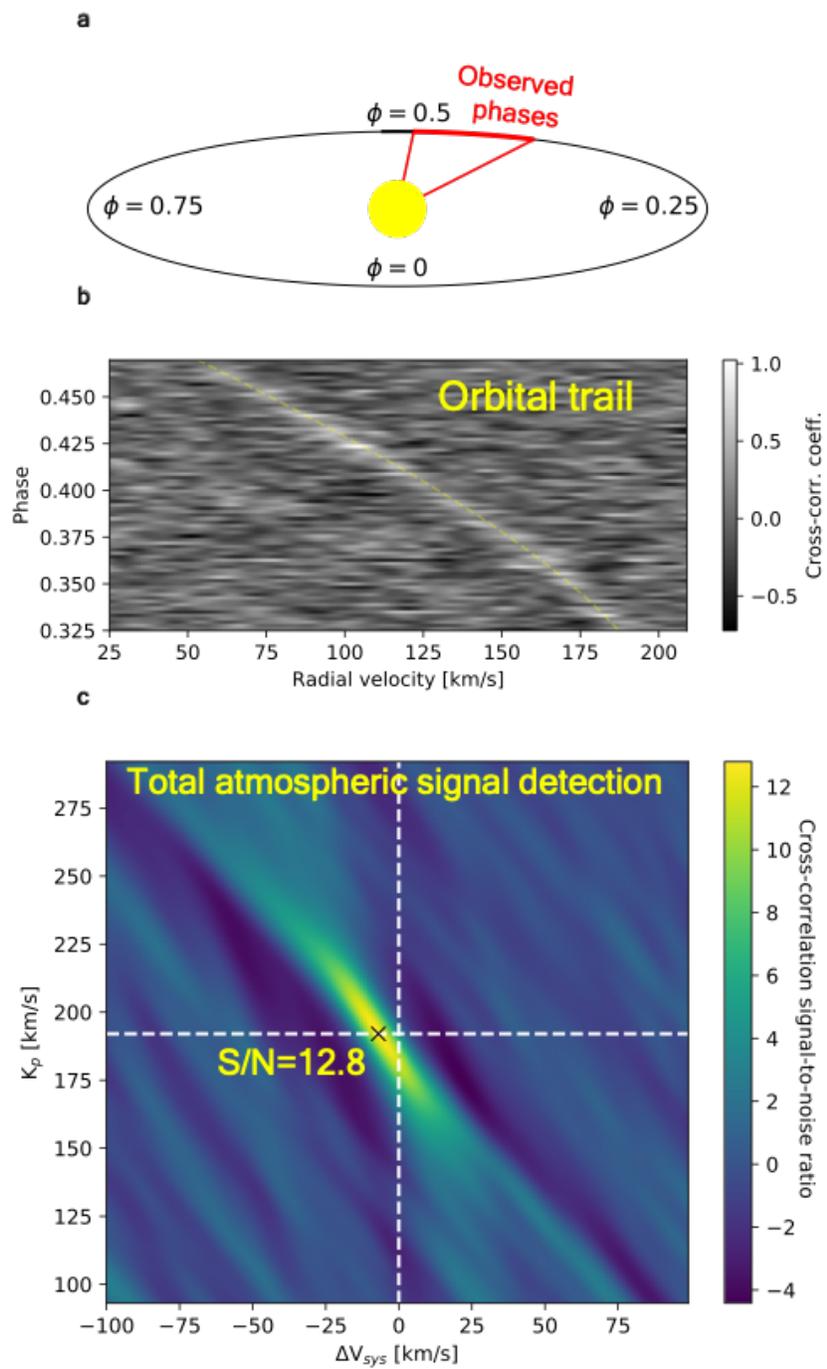

**Fig 1:** Summary of the planetary atmosphere signal detection. a) Illustration of the observed orbital phases (red), covering $0.32 < \varphi < 0.47$ (the orientation of the orbit is tilted for perspective). b) Cross correlation coefficient as a function of orbital phase/spectrum and planet velocity using a model template from the Bayesian inference procedure described in the text. The white trail corresponds to higher cross-correlation values (hence, atmospheric signal) and is consistent with the predicted velocity trail given the planetary orbital velocity and system velocity (light yellow dashed line). c) Atmospheric day-side thermal flux detection signal-to-noise (detection of absorption due to $H_2O$ and CO, see text) as a function of the planetary orbital velocity, $K_p$, and the relative system velocity ($\Delta V_{sys}$) (see Text). The significance was computed by subtracting off the mean of the cross-correlation map and then dividing through by the standard deviation of a box far from the planet velocity pair. White dashed lines indicate the known[8] velocities ($K_p$=192.06, $\Delta V_{sys}$=0 km s$^{-1}$) and the "×" denotes the location of the peak signal (S/N=12.8)

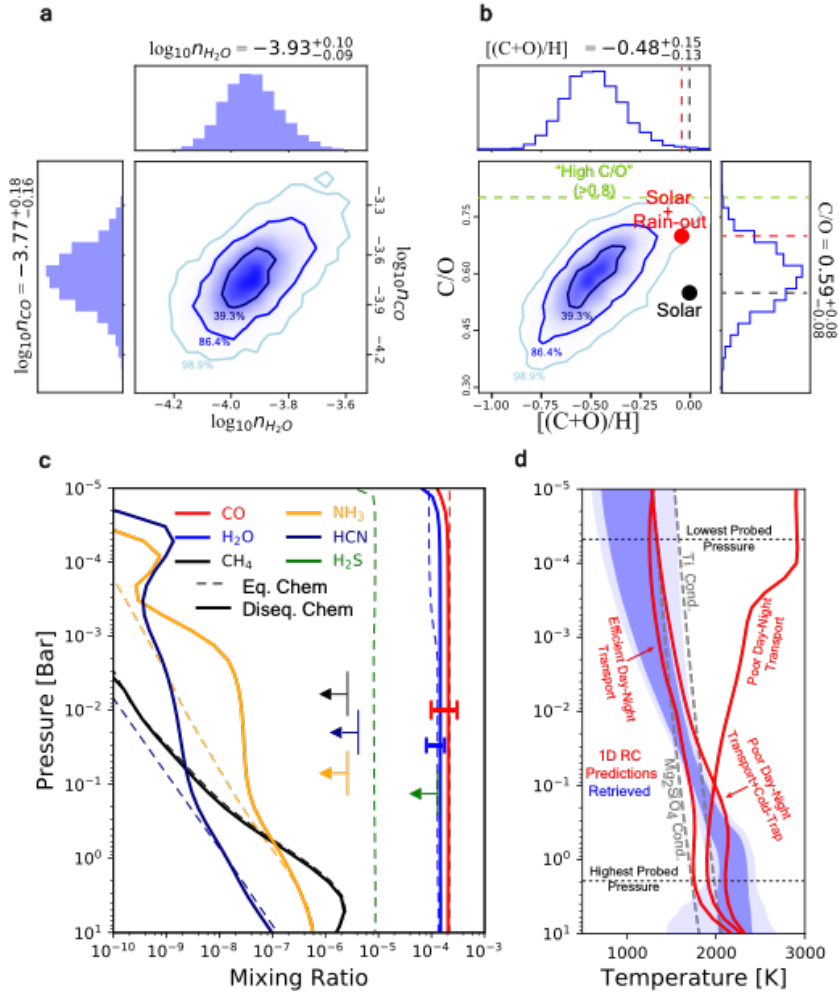

Fig 2: Summary of the composition and vertical thermal structure constraints, compared to predictions. a) marginalized and joint probability constraints for the $\log_{10}$ volume mixing ratios ($n$) of $H_2O$ and CO. b) marginalized and joint probability constraints for the atmospheric C/O and metallicity proxy, [(C+O)/H]. The solar abundance value[17] is given as the black point and the solar abundance value accounting for oxygen sequestration due to potential condensate rain out[24] on the night side[16] is shown as the red point. The 1(39.3%)- 2(86.4%)- and 3(98.9%)$\sigma$ joint probability contours are indicated in both a and b and the numerical values above each histogram are the marginalized median and 68% confidence interval range. c) Vertical abundance profiles for the major species predicted with both equilibrium (dashed) and disequilibrium (vertical transport and photochemistry, solid) chemistry (see Methods). d) Retrieved vertical temperature structure (magenta, 68 and 95% confidence intervals) compared to 1D radiative-convective equilibrium models with the coldest resulting from efficient day-to-night heat transport, the hottest poor heat transport, and the middle, poor heat transport but with nightside condensation of refractory species (see Methods).

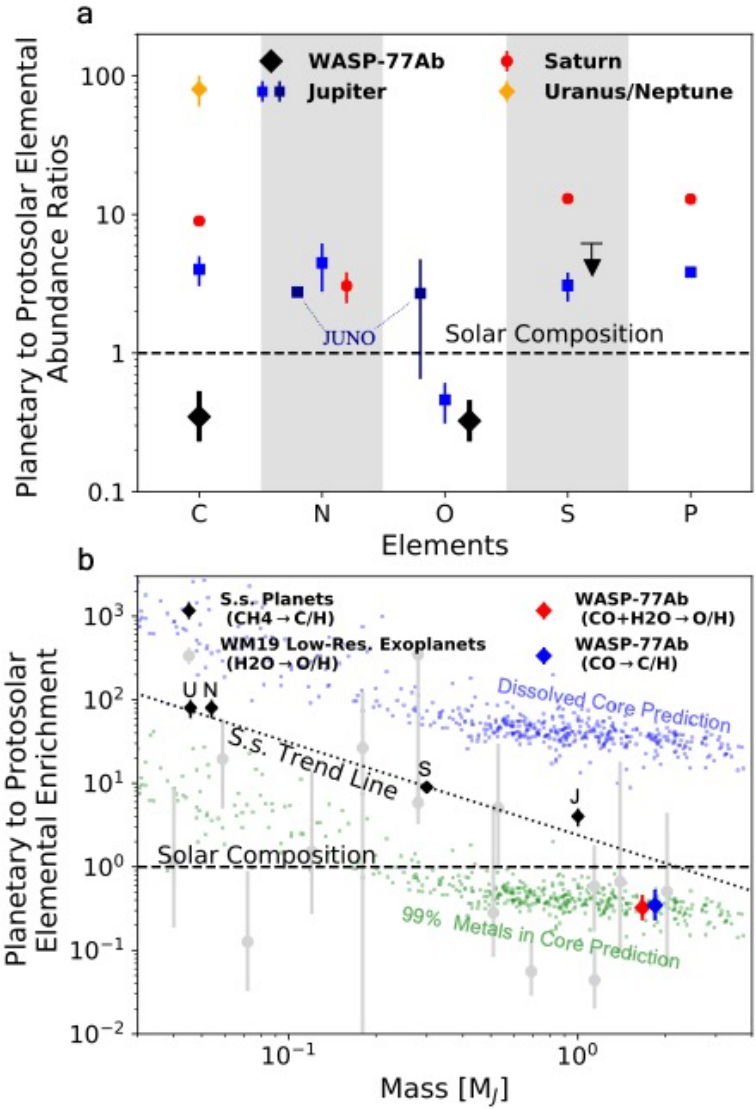

**Fig 3:** Comparison of the IGRINS WASP-77Ab abundance constraints with the solar system planets, exoplanets, and several predictions. a) elemental abundance constraints compared to the solar system planets (adapted from[25]) b) WASP-77Ab abundance constraints in the context of the mass vs. metallicity trend established by the solar system planets and populated with $H_2O$ based metallicity measurements with HST. The gray points are from a uniform HST transmission spectrum water retrieval analysis by Ref.[4]. The solar system planets (black diamonds) follow a decreasing trend (dotted line) with increasing mass[4,22]. The light blue and green dots are the predicted envelope enrichments for the gas rich planet population based upon their mass and radius measurements[26]. The blue dots assume a planet without a core with all metals (e.g., C and O) uniformly mixed throughout the gas, whereas the green dots assume that 99% of the metals are in a solid planetary core (1% in the envelope). The C and O elemental abundance constraints for WASP-77Ab are derived from the CO and $H_2O$ gas mixing ration constraints described in the main text.

# Methods

## Observations & Data Analysis

We observed a 4.7 hour continuous pre-eclipse sequence of WASP-77Ab with IGRINS on Gemini South as part of program GS-2020B-Q-249 on Dec. 14 2020. A sequence of of 79 A-B pairs (70 s each, or 140 s total per nod pair) covering a phase range ($\varphi$) between 0.325-0.47 achieving a per resolution-element signal-to-noise ratio of >200 on average (ED Fig. 1). The WASP-77 system is composed of a primary, WASP-77A (G8V $T_{eff}$=5500±80K, $M_K$=8.4) and a secondary WASP77B (K5V, $T_{eff}$=4700±200K, $M_K \sim$ 9.4), separated by 3.3". To avoid contamination we adjusted the slit (0.3") position angle (150°) on WASP-77A to maximize the separation perpendicular to the slit (~10 slit widths away).

The IGRINS Pipeline Package (PLP)[10,27] is used to reduce, optimally extract the spectra, and perform wavelength calibrations. A further wavelength calibration fine adjustment is made by applying a linear stretch re-alignment of each spectrum with the spectrum at the end of the sequence. This ensures self-consistent alignment to enable more robust telluric detrending. Finally, due to heavy telluric contamination (median atmospheric transmittance <0.7) we discard 11/54 orders near the edges of the H and K bands.

We use the Principle Component Analysis (PCA) method (singular value decomposition- SVD-with python's *numpy.linalg.svd*) of telluric detrending as it requires little hand tuning and has worked well on other instruments[11,28,29]. The SVD is applied directly to each individual order by zeroing out the first $N_{comp}$ eigenvalues that correspond to the dominant common mode in-time (left singular values) components followed by a reconstruction of the "telluric free" data matrix. We remove 4 PC's/SV's for all orders. We experiment with between 2 and 14 and found little difference for values from 4 - 10 (below).

## Modeling & Bayesian Inference Scheme

Rather than apply the standard cross-correlation analysis, we opt to use a Bayesian analysis/loglikelihood and modeling framework[7] so as to directly determine the atmospheric temperature/ abundance constraints. We update our radiative transfer model[7,30,31] to include the latest line lists from EXOMOL[32] and HITEMP[33] -CH4:HITEMP[34], CO:HITEMP[35], NH3:aCeTY super-lines, H2S:AYT2[36], HCN: Harris[37], along with H$_2$-H$_2$/He collision induced absorption: HITRAN[38]. We include separately the $^{12}C^{16}O$ and $^{13}C^{16}O$ lines weighted by the built-in terrestrial ratio of 1:89 ($^{13}C$:$^{12}C$) in the HITRAN/HITEMP line lists. When we refer to "CO" constraints in the main text, we refer to the main isotopologue, $^{12}C^{16}O$. Cross-sections are **pre-computed on a T-P grid (2.5 ≤ log$_{10}$(P) [bar] ≤ −6, 0.5 dex increments, 500 ≤ T ≤ 3000, 100 K increments)** with the *HELIOS-K* package[39,40] at 0.001 cm$^{-1}$ resolution and a Voigt wing cut of 100 cm$^{-1}$ and then interpolated down to a constant R=250K. For water we use the POKOZATEL[41] line list with methods/broadening parameters described in[42]. ED Fig. 2 summarizes the opacity sources at a representative temperature/pressure.

The 1D atmosphere is parameterized with constant-with-altitude volume mixing ratios for the aforementioned gases (6 gases with He+$H_2$ as the remainder with $n_{He}/n_{H_2}$=0.176)and the 6-parameter temperature profile scheme described in Ref[43] where:

$$T(P) = \begin{cases} \frac{1}{\alpha_1} = ln^2\left(\frac{P}{P_0}\right) + T_0 & P_0 \leq P < P_1 \\ \frac{1}{\alpha_2} = ln^2\left(\frac{P}{P_2}\right) + T_2 & P_1 \leq P < P_3 \\ T = T_3 & P \geq P_3 \end{cases}$$

where $T_2$ and $T_3$ are determined via continuity at $P_1$ and $P_3$, respectively.

For completeness we also include as a free parameter the CO isotopic abundance ratio, $^{13}C^{16}O/^{12}C^{16}O$ ($log_{10}$ relative to the terrestrial value of 1:89, more on this below). For HRCCS specific retrievals, we must also include the planet Keplerian and system velocities ($\Delta K_p$ and $\Delta V_{sys}$ relative to the literature[8] reported values of 192±11 and 1.6845±0.0004 km s$^{-1}$, respectively). Finally, we include as nuisance parameters a stretching term to the planet flux to account for uncertainties in the reported planet/star radius or data reduction induced stretching and a phase offset term to account for errors in the reported ephemera. We do not explicitly include a cloud in our forward model, however, the effect of an opaque gray cloud (e.g., a large optical depth set at a prescribed "cloud-top-pressure"[6,7]) can be mimicked with the $log_{10}(P_3)$ parameter in the above temperature profile parameterization. An isothermal slab would produce a similar spectral effect (a blackbody produced at $P_3$) as an opaque gray cloud-top pressure. If there were such a cloud present at higher altitudes, we would have retrieved a lower value of $P_3$ than the current upper limit of ~0.4 (ED Figure 3, $log_{10}(P_3)$ panel). Thus, we expect little to no impact on the retrieved temperature or gas abundances. Extended data Table 1 summarizes the parameters and their uniform prior ranges.

The model planet spectrum is convolved with both a planetary rotational ($v\sin(i)$=4.52 km/s) and an instrumental broadening (assumed to be Gaussian) kernel followed by an interpolation (using the *python scipy.interpolate.splrep/splev* functions (3$^{rd}$ order) to the data wavelength grid (accounting for the appropriate Doppler shift), and finally divided by a stellar model (either a PHOENIX model or a blackbody) and scaled by the planet-to-star area ratio[8].

Bayesian inference and model selection are performed using *pymultinest*[44,45] to evaluate the log-likelihood function[7]. The likelihood evaluation steps are the same as described in ref.[7] except that we use the PCA instead of the airmass detrending method. To do so, we save the $N_{comp}$ discarded eigenvectors from the SVD to reconstruct the telluric/systematic data matrix followed by a multiplicative injection of the model (Doppler shifted matrix of 1+($F_p/F_{star}$)($\lambda,\varphi$)). The PCA/SVD is then re-applied to the model injected data for each order. This model-injected matrix is then cross-correlated and corresponding log-likelihood evaluated and summed over each individual spectrum in each order against the true data matrix for each model/parameter instance. A typical retrieval under this set-up (17 parameters, 500 live points) runs in about 3 days (~270K likelihood evaluations) if utilizing 24 CPUs (for paralleling-pythons' joblib package– the model-injected data PCA computation for each order) and a single NVIDIA Tesla V100 GPU for the radiative transfer.

Table 1: Retrieved parameters and their prior range

| Parameter | Description | Prior Range |
|---|---|---|
| $\log_{10} n_i$ | log of the gas volume mixing ratios (i=$H_2O$, $CO$, $CH_4$, $H_2S$, $NH_3$, $HCN$) | -12 - 0 |
| [$^{12}CO/^{13}CO$] | $\log_{10}$ isotopic ratio relative to terrestrial (89:1) | -3 - 3 |
| $T_0$ | Top of Atmosphere Temperature [K] | 500-2000 |
| $\log_{10}(P_1)$ | layer 1-2 boundary pressure [bar] (see Fig. 1[43]) | -5.5 - 2.5 |
| $\log_{10}(P_2)$ | layer 2 knee pressure [bar] (see Fig. 1[43]) | -5.5 - 2.5 |
| $\log_{10}(P_3)$ | isothermal layer pressure start [bar] (see Fig. 1[43]) | -2 - 2 |
| $\alpha_1$ | upper atm. Temp. shape param. (see Eq. 2[43]) | 0.02 - 1 |
| $\alpha_2$ | middle atm. Temp. shape param. (see Eq. 2[43]) | 0.02 - 1 |
| d$K_p$ | Planet Orbital velocity relative to published [km/s] | -20 - 20 |
| d$V_{sys}$ | System velocity relative to published [km/s] | -20 - 20 |
| $\log_{10}(a)$ | $\log_{10}$ planet flux scale factor | -1 - 1 |
| d$\varphi$ | Orbital phase offset | -0.01 - 0.01 |

We note that we could have chosen other CCF-to-log-likelihood mappings such as those described in Ref.[46] and Ref.[47]. These mappings make different assumptions regarding the properties of the noise. Generally, however, in the presence of high signal to noise, resolution, and broad wavelength coverage observations, like those obtained with IGRINS, these assumptions are unlikely to have a strong impact on the retrieved parameter constraints (e.g, as shown in Refs.[7,47]).

## Extended Results

**Retrieved Constraints** ED Fig. 3 summarizes the full posterior/parameter constraints and representative best fit model (inset) and is from where the primary results (e.g., main text Fig. 2) discussed in the main text are derived. The temperature profile confidence intervals presented in main text Fig. 2d are reconstructed from random posterior draws[48,49]. The water and carbon-monoxide abundances are much more tightly constrained compared to those obtained with low resolution space-based observations with HST/Spitzer. Seemingly counter-intuitive, due to removal of the planetary continuum during the telluric removal process (PCA), absolute abundance constraints for select species have been routinely achievable[5,6,7,47] within these high-resolution specific Bayesian retrieval methods. This absolute (temperature-gradient and abundance degeneracy can be broken) abundance information can be extracted due to the different non-linear dependencies of absorption feature strengths on temperature and abundance.

There is a non-negligible offset (~1.5$\sigma$, see d$V_{sys}$ histogram in ED Fig. 3) in the relative system velocity. Such offsets are not uncommon[6,50,51] and can arise for a variety of reasons including uncertainties in the propagated mid-point timing during the event (e.g., the propagated eclipse midpoint uncertainty is ~260s which would correspond to a ~2.7 km/s velocity uncertainty at the observed orbital phases), a small previously unnoticed eccentricity, or perhaps, more intriguingly a combination of rotation (~4.5 km/s) which might preferentially blue shift a dayside hot spot and/or longitudinal temperature advection (west-to-east winds, ~2 km/s[52]). Whatever the source of the velocity offset, it is inconsequential for the chemical and thermal profile constraints as it is non-degenerate with those atmospheric parameters.

For legacy with past works we also include the "classic"[12,13,29,50] CCF analysis (ED Fig. 4) about the maximum likelihood solution summarized with individual gas "detections" in the $K_p$-$\Delta V_{sys}$ plane and slices along the systemic velocity axis at the literature reported $K_p$ value. We also include, for comparison purposes, these same data products in delta-log-likelihood space (right column of ED Fig. 4). It is worth noting how the log-likelihood mapping boosts the signal of CO by adding information about the line shape and amplitude relative to the continuum, which is what ultimately enable absolute abundance constraints.

**Physical Plausibility Assessment** We assess the physical/chemical plausibility of the retrieved quantities with a 1D radiative-convective-thermochemical-equilibrium[53,54,55,56] and a chemical kinetics -transport-photochemical solver[57]. These tools self-consistently predict the 1D temperature profile and molecular abundances given the incident stellar flux (or scaling) and elemental abundance inventory ([M/H]=-0.4, C/O=0.58).

The results of this exercise are shown in main text Figs. 2c,d. We explore several plausible chemical/radiative scenarios: (i) "efficient day-night transport" which permits the planet to evenly re-radiate over both the day-and-night hemispheres[58], (ii) "poor day-night transport", in which the planet only re-radiates over the dayside hemisphere[58], (iii) "poor day-night transport+cold trap" which is the same as (ii) but removes UV-optical absorbing refractory opacities (TiO, VO, FeH, CrH, MgH, etc.), to mimic loss due to nightside condensation and what would be nominally predicted[59,60], and (iv) thermochemical equilibrium ("Eq. Chem") vs. photochemical-transport kinetics ("Diseq. Chem"). The latter (chemistry in main text Fig. 2c) assumes the temperature profile from (i) for simplicity. The retrieved molecular abundances are consistent with the plausible chemistry (photochemistry/transport matter little as this is a hot planet). The retrieved temperature profile is consistent with both the cooler two temperature profile scenarios, suggesting either daynight-cold trapping and/or efficient day-to-night heat transport.

**Elemental Abundance Determinations & Interpretation** The chemical plausibility, dominance of CO and $H_2O$, and uniformity with pressure/altitude permits us to directly compute the C and O enrichments and carbon-to-oxygen ratio (main text Fig. 1b). The total C abundance is given by CO/H, O by ($H_2O$+CO)/H, and H by $2H_2$ (where H2=0.837- the equilibrium chemistry value at WASP-77Ab temperatures). Solar abundances from[17] (C/H=2.95×10$^{-4}$, O/H =5.37×10$^{-4}$, C/O=0.55) are used when referencing the "relative to solar values" (e.g., main text Fig. 3) with "[]" referring to the $\log_{10}$ relative to solar. This results in a [C/H]=-0.46±0.17 (0.24 - 0.51×Solar), [O/H]=-0.49±0.13, (0.24 - 0.44×Solar) , [(C+O)/H]=-0.48±0.14 (0.24 - 0.46×Solar), and a C/O= 0.59±0.08. [(C+O)/H] is used as a proxy for the total metal enrichment ([M/H]). We include as a reference in main text Fig. 2b the solar values (the black solar point) and the rain-out value (red point) whereby O is lost into refractory condensates (possibly on the nightside, assuming 3.28 O atoms per Si atom from silicate cloud formation[24] results in a 22% reduction in O). If we "correct" for the loss of O due to condensate formation, then we obtain a [(C+O)/H]=-0.41±0.14 (0.29 - 0.54×Solar) and C/O=0.46±0.08. We use the "non-rainout" abundances in the main-text discussion and Fig. 3.

The C and O abundances for WASP-77Ab are interpreted through the lens of the solar system abundance determinations, the representative exoplanet population abundances as measured with low spectral resolution platforms (e.g., HST), and theoretical models in main text Fig. 3. To compare to the solar system (main text Fig. 3b) we use the abundances given in Table 1 in Ref. [25] (from references therein), with recent updates from JUNO[23]. It is worth noting, however, that the reported JUNO value for O (based upon $H_2O$ via the microwave radiometer equatorial measurements), while seemingly greater than the Galileo Probe "hot spot" measurement, is technically not a bounded constraint, rather more of an upper limit (see Fig. 5b in Ref.[23]) cannot rule out zero abundance alone.[23] Comparatively, with a single observation, our measurements provide bounded constraints on both C and O at sub-solar/stellar values.

Main text Fig. 3b places WASP-77Ab's abundance determinations in the context of low resolution HST $H_2O$/O measurements[4] (gray points), the solar system $CH_4$/C-based measurements (from Fig 3a) and trend line, and interior structure-based envelope metallicity predictions[26] (blue, green dots) as a function of planetary mass. If all abundances scaled proportionally to the total envelope metallicity, and the population synthesis predictions from Ref. [63] and Ref. [1] were true, we would expect exoplanet atmosphere metal enrichment's to loosely follow the solar system trend line (dotted). There is clearly no trend with the $H_2O$ based O measurements in[4] (see also[63,64]), though the uncertainties are quite large. This could be suggestive that perhaps O is "depleted" (e.g., via high C/O), though without a C measurement for each planet this cannot be confirmed. The C and O abundances in WASP-77Ab both fall well below the trend line, and even below solar composition.

Ref.[26] provide predictive models for the maximum metal enrichment (based upon O) for the exoplanet population given their measured mass and radius for a "core-less" planet, e.g., the metals and gas are well mixed throughout the entire planet. This is clearly extreme as these values (blue dots, main text Fig. 3b) vastly overshoot the measured Jupiter and Saturn envelope enrichment's, suggesting a large fraction of metals must be sequestered into a solid core (on the order of 90%). To match the retrieved depletion for WASP-77Ab, approximately 99% of the accreted metals must be in the planetary core (assuming the observed atmospheric composition is representative of the entire envelope). One cannot tell the formation story with a single planet as the vast complexities between the composition of accreted gas and the partitioning of metals/solids within the core are not yet cleanly predictable. A larger survey of planets with precisions obtained here could shed light on this seemingly insurmountable problem.

### A Series of Robustness Tests

To test the robustness of the abundance and temperature profile constraints, we perform a battery of tests that explore the impact of data processing and modeling assumptions on the retrieved $H_2O$, CO abundances, and temperature profile (summarized in ED Fig. 5).

The first test is used to evaluate the influence of the TP-profile parameterization ("Temperature Profile Parameterization", top histogram row, first TP-profile panel). The atmospheric parameterization is identical to that described above, but replacing the parameterization from Ref.[43] with the 3-parameter analytic prescription from Refs.[66,67]. This has virtually no effect on the retrieved gas abundances, and a slight change in the temperature gradient in the ~1 - 0.01 bar region.

The second test gauges the impact of spatial heterogeneity's in temperature[52,67]. If the planet had strong spatially varying temperatures, like a dominating hot spot, we would expect to retrieve different temperatures between the first half and the second half of the observing sequence as the heterogeneities rotate into/out of view. To test this, we broke the observing sequence in half (compared to the full sequence) as if each were its own separate observation, each having 40 (39) frames/spectra per sequence. The entire PCA analysis/retrieval procedure above was then applied to each half-sequence. For computational reasons we used the faster 3-parameter analytic TP-profile prescription from Refs.[65,66] (since this choice did not matter in the first test). These results are shown ("3D Temperature Effects") in the middle histogram row and middle TP-profile panel of ED Fig. 5. Again, this had little influence other than increasing the uncertainties on the abundances and TP-profile due to the reduced data set size per half-sequence. This suggests that a "1D" atmosphere/retrieval is sufficient in this case and does not result in any measurable bias.

The last sequence of tests explores common processing/model assumptions ("Processing Assumptions", bottom histogram row, last TP-profile panel). The Reference (REF) model here assumes the 3-parameter analytic TP-profile prescription from Refs.[66,67], $H_2O$ and CO as the only abundance free parameters (as we only obtained upper limits on the others above), a PHOENIX stellar spectrum (for $F_p/F_{star}$), a Gaussian instrumental profile consistent with R=45K, no rotational broadening (it too does not matter), 4-principal components in the processing, and R=250K model resolution (as in the main analysis). We explored these dimensions/assumptions one-at-a-time by 1) changing the instrumental profile to that of an R=71K instrument (IP/71K, narrower), 2) using a blackbody stellar spectrum (BB Star) instead of a PHOENIX model, 3) 8-principal components removed in the PCA (8 PC), and R=500K model resolution (R=500K xsecs). None of these assumptions had a significant influence on the retrieved $H_2O$ and CO abundances or temperature profile.

Finally, we undergo an independent Bayesian/retrieval analysis using an entirely independent tool/code (HyDRA-H[5,68]), but also utilizing the log-likelihood mapping from[7] and PCA for airmass detrending. A comparison of a subset of common parameters is shown in ED Fig. 6. In this comparison, HyDRA-H retrieves for identically the same parameters as described in ED Table 1 with the following differences: the $NH_3$, HCN, $H_2S$, and $CH_4$ gas mixing ratios are not included, and no orbital phase offset parameter is included. The results sufficiently agree, with only a slight (1.5 $\sigma$) offset in the median values of the retrieved CO abundance and small differences in the slope of the TP-profile. We also note (not shown) that the retrieved velocities and scale factor are in very good agreement as well. These differences don't affect our main conclusions that the overall C and O enrichment is low, the C/O constraints rule out high C/O scenarios, and the temperature profile decreases with decreasing pressure (e.g., no thermal inversion).

We thus conclude that the resulting constraints, and subsequent derivatives thereof, presented in the main text are resilient against the common data analysis choices and modeling assumptions.

## A Potential $^{13}C^{16}O/^{12}C^{16}O$ Constraint

Isotopic abundance ratios provide an additional composition dimension[25,69,70] with which we can explore planet formation and atmospheric chemistry due to their sensitive mass/temperature dependent fractionation. High resolution observations are potentially sensitive[71] to isotopic abundance ratios for select molecules, specifically, for IGRINS, those of CO-$^{12}C^{16}O/^{13}C^{16}O$ (primarily near the 2.3 $\mu$m CO bandhead). For these reasons, we include this ratio as a free parameter (ED Table 1). Surprisingly (ED Fig. 3), we obtain a bounded constraint with [$^{12}C^{16}O/^{13}C^{16}O$]=$-0.65^{+0.33}_{-0.29}$ (0.11-0.48×terrestrial, or $^{12}C^{16}O/^{13}C^{16}O$ of 10.2 - 42.6 ). We do not see any signature within the CCF itself, though this is not unexpected[72].

To bolster confidence in the isotopic ratio constraint, we perform a reverse injection and retrieval test. To do this, we first perform a simplified retrieval (the 3-parameter TP-profile from §4, H$_2$O, CO, CO isotope ratio, the velocities, and stretch factor) on the data. We then reverse inject[6] (via division) the maximum likelihood spectrum (1+$F_p/F_{star}$, appropriately convolved and Doppler shifted to each frame/phase) into the raw data sequence to remove the nominal planetary signal. Into this "best fit removed" data set we then re-inject the best fit model spectrum (through multiplication) but with the $^{13}$C$^{16}$O abundance set to zero. We then re-retrieve on this model injected dataset, resulting in only an upper limit on [$^{13}$C$^{16}$O/$^{12}$C$^{16}$O], as expected (ED Fig. 6, top panels, black histograms). This suggests that there is real information in the data producing this constraint, that may not necessarily result in strong detection's in the classic sense.

Finally, in ED Fig. 6, bottom panel, we compare the WASP-77Ab $^{13}$C/$^{12}$C constraint (via CO) to common solar system bodies and various reference values. It is currently beyond the scope of this manuscript to speculate as to why WASP-77Ab has a notably lower ratio than (enhanced $^{13}$C) compared to solar system, suffice it to say that protoplanetary disk chemistry models can produce a broad range of $^{13}$C/$^{12}$C in CO as a function of mid-plane height and radial distance from the star[69]. We purposefully choose not to strongly emphasize isotopic abundance constraint result in the main-text as more work needs to be done within the community to determine how to reliably quantify isotopic measurements-e.g., what is a detection?-this is non-trivial for these types of observations[71]. Future observations are needed both for this planet and for others in order to determine the commonality of such constraints.

## Methods References

27. Lee, Jae-Joon & Gullikson, Kevin. PLP: v2.1 alpha 3 [Data set]. Zenodo. http://doi.org/10.5281/zenodo.56067 (2016)

28. Piskorz, D. *et al.* Evidence for the Direct Detection of the Thermal Spectrum of the Non-Transiting Hot Gas Giant HD 88133 b. *Astrophys. J.* **832**, 131 (2016).

29. Giacobbe, P. *et al.* Five carbon- and nitrogen-bearing species in a hot giant planet's atmosphere. *Nature.* **592**, 205-208 (2021).

30. Line, M. R. *et al.* A Systematic Retrieval Analysis of Secondary Eclipse Spectra. I. A Comparison of Atmospheric Retrieval Techniques. *Astrophys. J.* **775**, 137 (2013).

31. Line, M. R. *et al.* Uniform Atmospheric Retrieval Analysis of Ultracool Dwarfs. II. Properties of 11 T dwarfs. *Astrophys. J.* **848**, 83 (2017).

32. Tennyson, J. *et al.* The 2020 release of the ExoMol database: molecular line lists for exoplanet and other hot atmospheres. *J. Quant. Spectrosc. Radiat. Transf.* **255**, 107228 (2020).

33. Rothman, L. S. *et al.* HITEMP, the high-temperature molecular spectroscopic database. *J. Quant. Spectrosc. Radiat. Transf.* **111**, 2139-2150 (2010).

34. Hargreaves, R. J. *et al.* An Accurate, Extensive, and Practical Line List of Methane for the HITEMP Database. *Astrophys. J. Suppl. Ser.* **247**, 55 (2020).

35. Li, G. *et al.* Rovibrational Line Lists for Nine Isotopologues of the CO Molecule in the X $^1\Sigma^+$ Ground Electronic State. *Astrophys. J. Suppl. Ser.* **216**, 15 (2015).

36. Azzam, A. A. A., Tennyson, J., Yurchenko, S. N. & Naumenko, O. V. ExoMol molecular line lists - XVI. The rotation-vibration spectrum of hot $H_2S$. *Mon. Not. R. Astron. Soc.* **460**, 4063-4074 (2016).

37. Barber, R. J. *et al.* ExoMol line lists - III. An improved hot rotation-vibration line list for HCN and HNC. *Mon. Not. R. Astron. Soc.* **437**, 1828-1835 (2014).

38. Karman, T. *et al.* Update of the HITRAN collision-induced absorption section. *Icarus* **328,** 160-175 (2019).

39. Grimm, S. L. & Heng, K. HELIOS-K: An Ultrafast, Open-source Opacity Calculator for Radiative Transfer. *Astrophys. J.* **808**, 182 (2015).

40. Grimm, S. L. *et al.* HELIOS-K 2.0 Opacity Calculator and Open-source Opacity Database for Exoplanetary Atmospheres. *Astrophys. J. Suppl.* **253**, 30 (2021).

41. Polyansky, O. L. *et al.* ExoMol molecular line lists XXX: a complete high-accuracy line list for water. *Mon. Not. R. Astron. Soc.* **480**, 2597-2608 (2018).

42. Gharib-Nezhad, E. *et al.* EXOPLINES: Molecular Absorption Cross-Section Database for Brown Dwarf and Giant Exoplanet Atmospheres. *Astrophys. J. Suppl. Ser. .* **254**, 34 (2021)

43. Madhusudhan, N. & Seager, S. A Temperature and Abundance Retrieval Method for Exoplanet Atmospheres. *Astrophys. J.* **707**, 24-39 (2009).

44. Buchner, J. *et al.* X-ray spectral modelling of the AGN obscuring region in the CDFS: Bayesian model selection and catalogue. *Astron. & Astrophys.* **564**, A125 (2014)

45. Feroz, F., Hobson, M. P. & Bridges, M. MULTINEST: an efficient and robust Bayesian inference tool for cosmology and particle physics. *Mon. Not. R. Astron. Soc.* **398**, 1601-1614 (2009).

46. Zucker, S. Cross-correlation and maximum-likelihood analysis: a new approach to combiningcross-correlation functions. *Mon. Not. R. Astron. Soc.* **342**, 1291-1298 (2003).

47. Gibson, N. P. *et al.* Detection of Fe I in the atmosphere of the ultra-hot Jupiter WASP-121b, and a new likelihood-based approach for Doppler-resolved spectroscopy. *Mon. Not. R. Astron. Soc.* **493**, 2215-2228 (2020).

48. Line, M. R., Knutson, H., Wolf, A. S. & Yung, Y. L. A Systematic Retrieval Analysis of Secondary Eclipse Spectra. II. A Comparison of Atmospheric Retrieval Techniques. *Astrophys. J.* **775**, 137 (2013).

49. Line, M. R., Knutson, H., Wolf, A. S. & Yung, Y. L. A Systematic Retrieval Analysis ofSecondary Eclipse Spectra. II. A Uniform Analysis of Nine Planets and their C to O Ratios. *Astrophys. J.* **783**, 70 (2014).

50. Hoeijmakers, H. J. *et al.* Atomic iron and titanium in the atmosphere of the exoplanet KELT9b. *Nature* **560**, 453-455 (2018). 1808.05653.

51. Brogi, M. *et al.* Detection of Molecular Absorption in the Dayside of Exoplanet 51 Pegasi b? *Astrophys. J.* **767**, 27 (2013).

52. Beltz, H., Rauscher, E., Brogi, M. & Kempton, E. M. R. A Significant Increase in Detectionof High-resolution Emission Spectra Using a Three-dimensional Atmospheric Model of a Hot Jupiter. *Astron. J.* **161**, 1 (2021).

53. Piskorz, D. *et al.* Ground- and Space-based Detection of the Thermal Emission Spectrum of the Transiting Hot Jupiter KELT-2Ab. *Astron. J.* **156**, 133 (2018).

54. Gharib-Nezhad, E. & Line, M. R. The Influence of $H_2O$ Pressure Broadening in High-metallicity Exoplanet Atmospheres. *Astrophys. J.* **872**, 27 (2019).

55. Arcangeli, J. *et al.* $H^-$ Opacity and Water Dissociation in the Dayside Atmosphere of the Very Hot Gas Giant WASP-18b. *Astrophys. J. Lett.* **855**, L30 (2018).

56. Mansfield, M. *et al.* An HST/WFC3 Thermal Emission Spectrum of the Hot Jupiter HAT-P-7b. *Astron. J.* **156**, 10 (2018).

57. Tsai, S.-M. *et al.* Toward Consistent Modeling of Atmospheric Chemistry and Dynamics in Exoplanets: Validation and Generalization of the Chemical Relaxation Method. *Astrophys. J.* **862**, 31 (2018).

58. Fortney, J. J., Lodders, K., Marley, M. S. & Freedman, R. S. A Unified Theory for the Atmospheres of the Hot and Very Hot Jupiters: Two Classes of Irradiated Atmospheres. *Astrophys. J.* **678**, 1419-1435 (2008).

59. Perez-Becker, D. & Showman, A.P. Atmospheric heat redistribution on hot Jupiters. *Astrophys. J.* **776**, 134 (2013).

60. Parmentier, V., Showman, A. P. & Fortney, J. J. The cloudy shape of hot Jupiter thermal phase curves. *Mon. Not. R. Astron. Soc.* **501**, 78-108 (2021).

61. Burrows, A., Sudarsky, D. & Hubeny, I. Theory for the Secondary Eclipse Fluxes, Spectra,Atmospheres, and Light Curves of Transiting Extrasolar Giant Planets. *Astrophys. J.* **650**, 1140-1149 (2006).

62. Fortney, J. J. *et al.* A Framework for Characterizing the Atmospheres of Low-Mass Low Density Transiting Planets. *Astrophys. J.* **775**, 80 (2013).

63. Pinhas, A., Madhusudhan, N., Gandhi, S. & MacDonald, R. $H_2O$ abundances and cloud properties in ten hot giant exoplanets. *Mon. Not. R. Astron. Soc.* **482**, 1485-1498 (2019).

64. Fisher, C. & Heng, K. Retrieval analysis of 38 WFC3 transmission spectra and resolution ofthe normalization degeneracy. *Mon. Not. R. Astron. Soc.* **481**, 4698-4727 (2018).

65. Guillot, T. On the radiative equilibrium of irradiated planetary atmospheres. *Astron. Astrophys.* **520**, A27 (2010)

66. Line, M. R. *et al.* Information Content of Exoplanetary Transit Spectra: An Initial Look. *Astrophys. J.* **749**, 93 (2012).

67. Feng, Y. K. *et al.* The Impact of Non-uniform Thermal Structure on the Interpretation of Exoplanet Emission Spectra. *Astrophys. J.* **829**, 52 (2016).

68. Gandhi, S. *et al.* Molecular cross-sections for high-resolution spectroscopy of super-Earths, warm Neptunes, and hot Jupiters. **495**, 224-237 (2020).

69. Woods, P. M. & Willacy, K. Carbon Isotope Fractionation in Protoplanetary Disks. *Astrophys. J.* **693**, 1360-1378 (2009).

70. Marboeuf, U., Thiabaud, A., Alibert, Y. & Benz, W. Isotopic ratios D/H and $^{15}N/^{14}N$ in giant planets. *Mon. Not. R. Astron. Soc.* **475**, 2355-2362 (2018).

71. Molliere, P. & Snellen, I. A. G.` Detecting isotopologues in exoplanet atmospheres using ground-based high-dispersion spectroscopy. *Astron. & Astrophys.* **622**, A139 (2019).

72. Wood, P. L., Maxted, P. F. L., Smalley, B. & Iro, N. Transmission spectroscopy of the sodium 'D' doublet in WASP-17b with the VLT. *Mon. Not. R. Astron. Soc.* **412**, 2376-2382 (2011).

**Data Availability** The raw PLP extracted IGRINS data files and subsequent data products are available here:

https://www.dropbox.com/sh/0cxfolfmrs8ip37/AABZYoHr8nuRlHJG84dArX4ea?dl=0

**Code Availability** The IGRINS PLP used to perform the initial reduction and extraction by the instrument team is available at https://github.com/igrins/plp. The barycenter correction and planetary phase calculations were made using the python *astropy* library found here https://www.astropy.org/. Python *Numpy* specific tools are noted in the text (e.g., the SVD for the PCA). The chemical abundance analysis/physical plausibility assessment made use of the *VULCAN* chemical kinetics tool (https://github.com/exoclime/VULCAN). Absorption cross-sections were generated using the *HELIOS-K* tool (https://helios-k.readthedocs.io/en/latest/). Finally, we make available a an end-to-end python2/GPU HRCCS retrieval code example available here https://www.dropbox.com/sh/0cxfolfmrs8ip37/AABZYoHr8nuRlHJG84dArX4ea?dl=0 which makes use of the *pymultinest* nested-sampling package (https://johannesbuchner.github.io/PyMultiNest/), *joblib* loop parallelization package (https://joblib.readthedocs.io/en/latest/), and *corner.py* (https://corner.readthedocs.io/en/latest/)

## Extended Data Figures

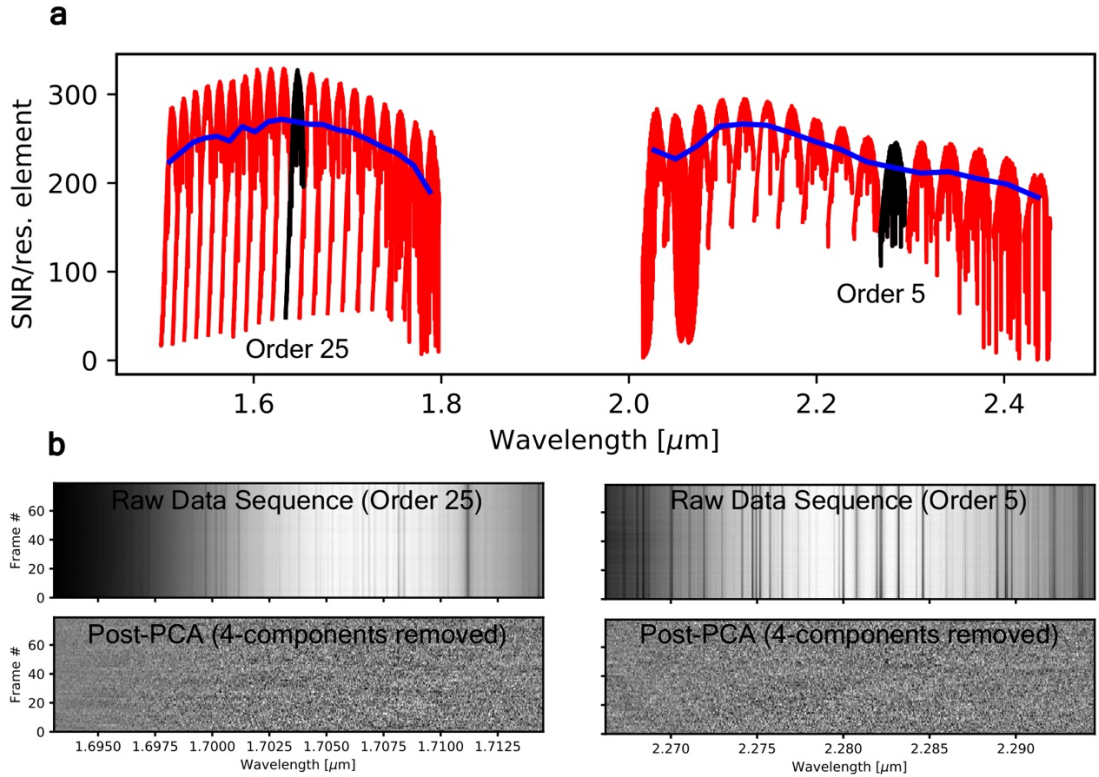

**ED Fig. 1**: Summary of the data and PCA procedure. a) shows the median per-resolution element signal-to-noise for each order for the night (in red). The blue curve is the median SNR in both time and over an individual order. b) shows example raw data cubes (top row)–spectra vs. time/frame for representative two orders (25, 5). Stationary tellurics show up as vertical dark streaks. Wavelength dependent gradient is due to the echelle blaze throughput. The PCA/SVD method can remove these stationary features, leaving behind the planetary signal buried in the noise (bottom row). We use these "Post-PCA" frames for the subsequent cross-correlation/retrieval analysis (repeated for all 43 use orders).

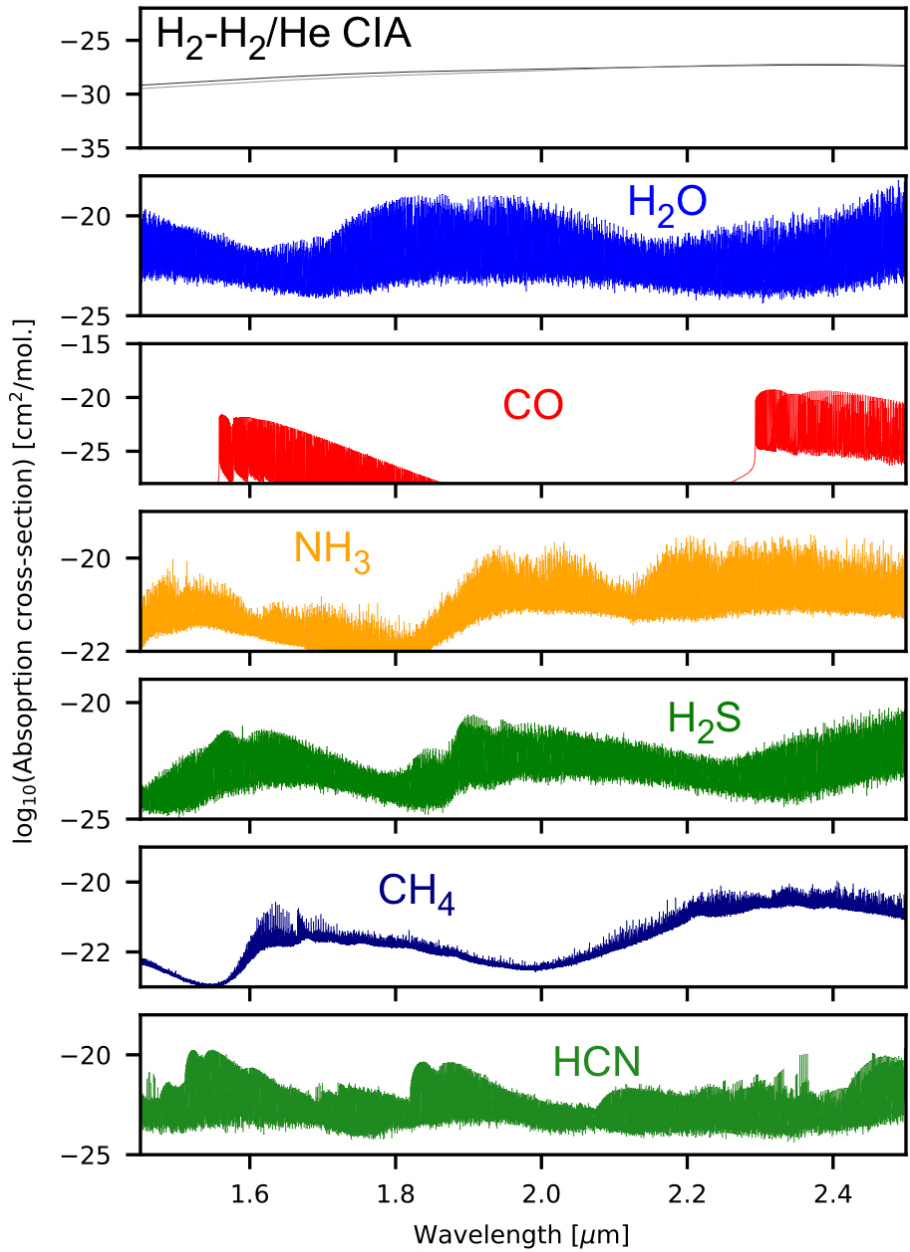

**ED Fig. 2**: Absorption cross-sections for the molecules considered in the retrieval analysis (for 0.01 bar, 1600K).

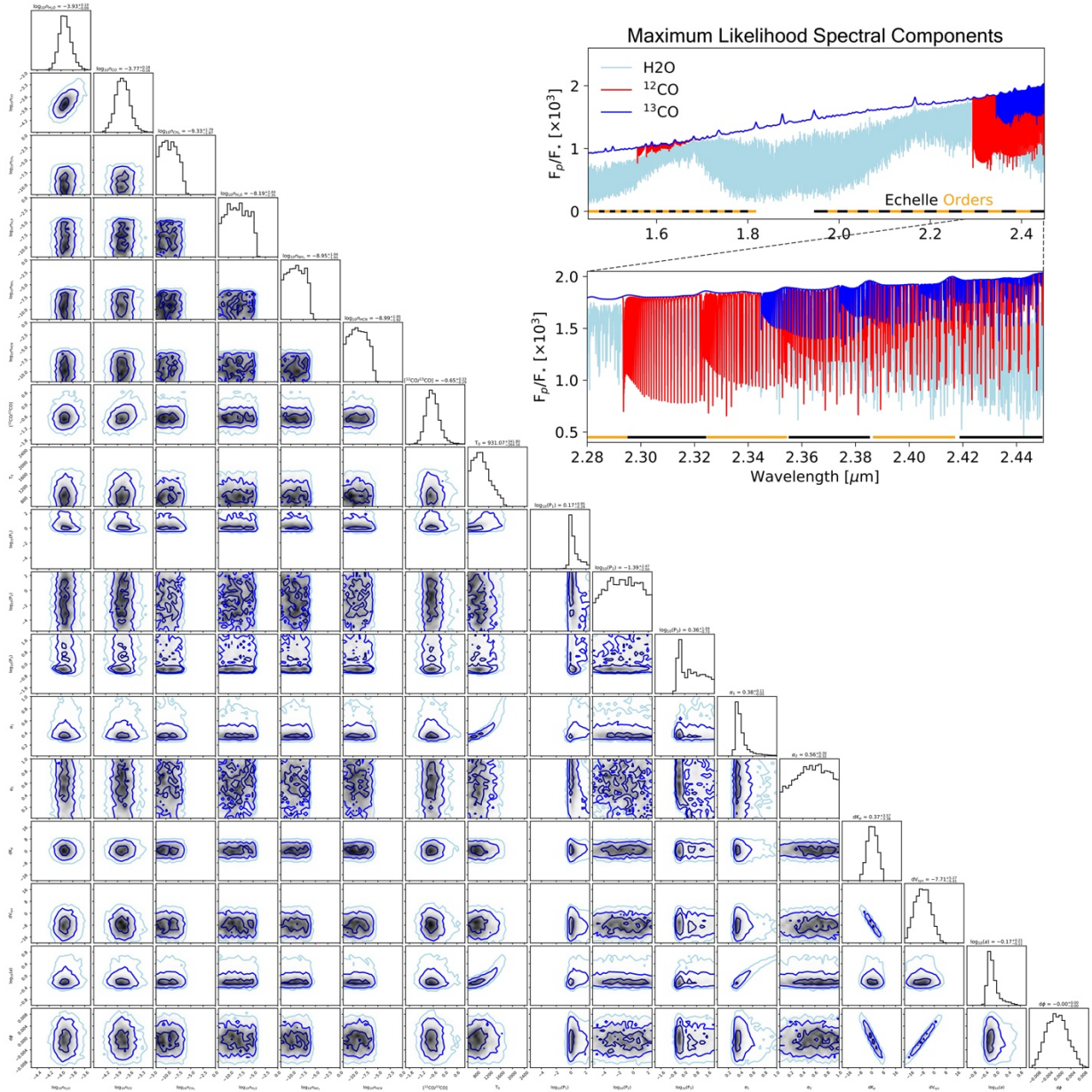

**ED Fig. 3**: Corner plot summary of the posterior probability distribution from the main-text retrieval analysis. Note the bounded constraints on water, CO, and the isotopic ratio, but upper limits only on the other species. Note, we retrieve [$^{13}C^{16}O/^{12}C^{16}O$] but plot the inverse,[$^{12}C^{16}O/^{13}C^{16}O$] to facilitate comparisons to literature reported values (in ED Fig. 6) The inset shows the molecular components of the maximum likelihood model spectrum. Figure generated with *corner.py*

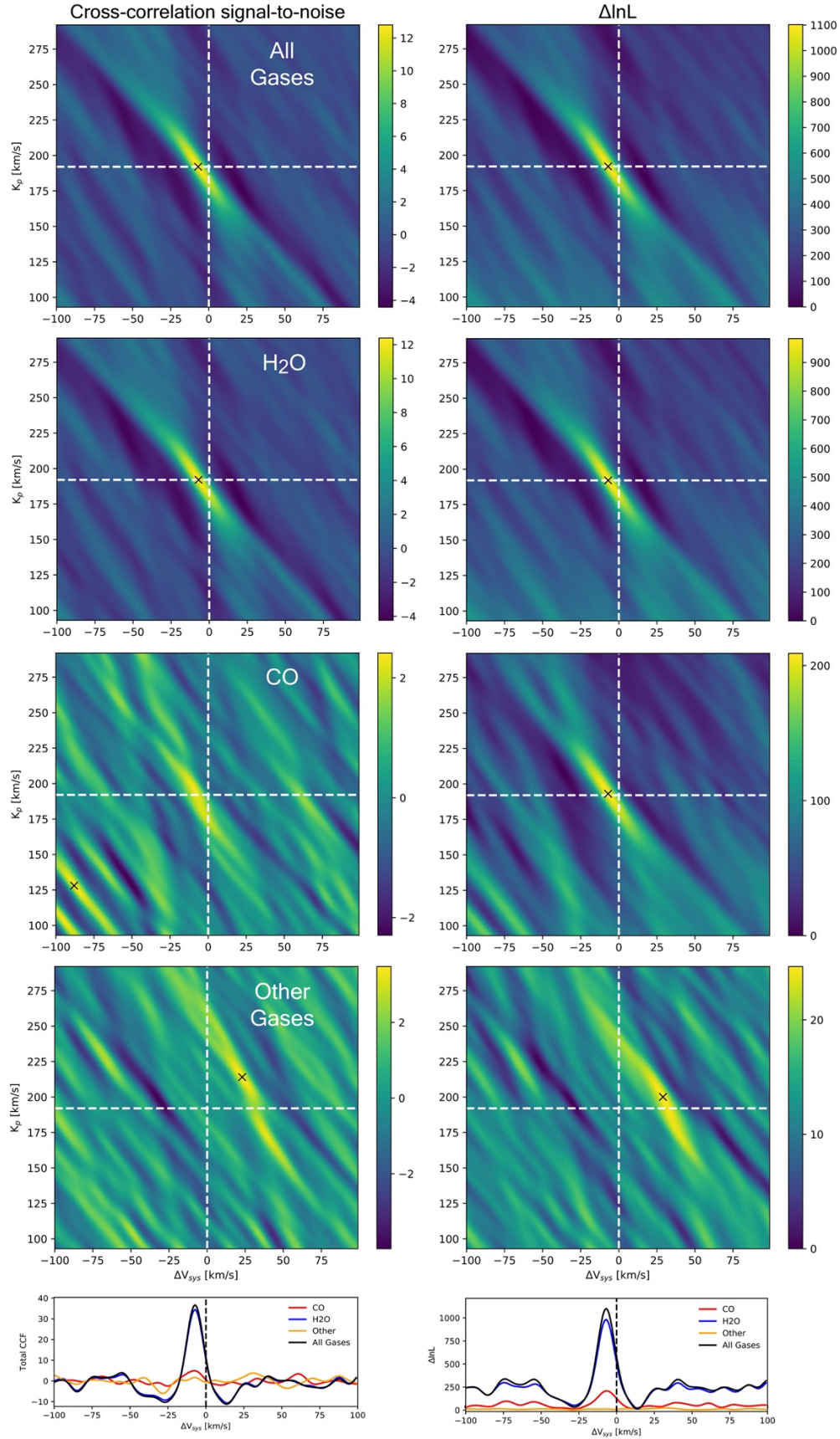

**ED Fig. 4**: Classic cross-correlation analysis data products. The model template used to in this cross-correlation analysis is the spectrum resulting from the maximum likelihood solution found by the retrieval analysis. The left column illustrates the gas detection's (all gases, $H_2O$, CO, and other–$NH_3$+$H_2S$+HCN+$CH_4$) in the standard $K_p$-$V_{sys}$ plane, with a slice in $V_{sys}$ along the literature reported $K_p$ at the bottom. The detection maps are constructed by subtracting the mean total CC, then dividing by an "off peak" (a boxed region in the lower left corner of each panel) CC standard deviation. Using this method, only $H_2O$ is strongly detected, with a hint of CO present at the expected velocities. The right column reproduces analogous products using the log-likelihood formalism[7] ($\Delta\log L$ relative to the minimum), resulting in a stronger presence of CO. We emphasize, that while such maps may be instructive for "detecting" species or "atmosphere", they do not marginalize over all of the degeneracy, nor do they maximize the information content in the data. This is why in our analysis, we focus on the the results arising from the more comprehensive log-likelihood/retrieval formalism.

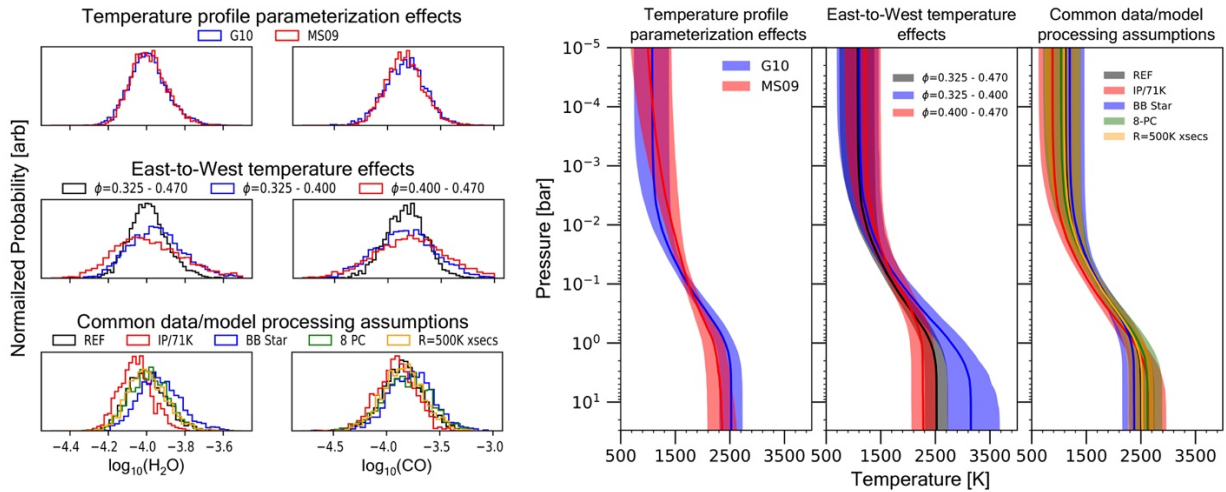

**ED Fig. 5**: Robustness test analyses summary using the $H_2O$, CO, and temperature profile constraints as the metrics for assumption impact. The top row of histograms and first TP-profile histogram demonstrate the lack of impact of TP-profile parameterization. The middle panel of histograms and middle TP-profile panel show that there is little impact due to any presence of temperature heterogeneities on the hemisphere(s) observed during the sequence. Finally, the bottom panel of histograms and last TP-profile panel illustrate the lack of impact of various data analysis and other minor modeling assumptions. In short, the retrieved abundances and temperature profile constraints are largely resilient against most common assumptions.

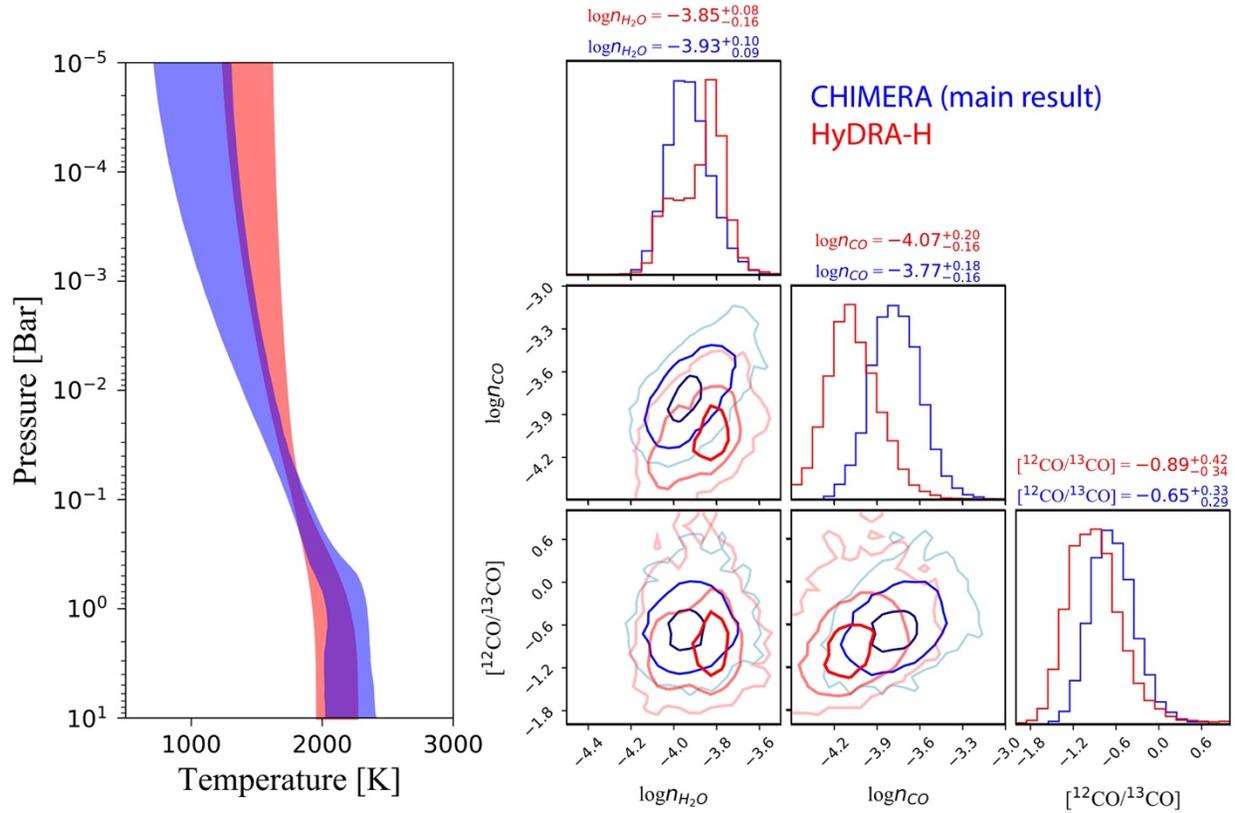

**ED Fig. 6**: Bayesian inference/retrieval tool comparison on the IGRINS data. The temperature profiles are compared in the left most panel and a subset of the abundances in the corner plot on the right. Each model uses slightly different atmospheric parameterization assumptions with the core radiative transfer aspects (solver, opacities) independently developed.

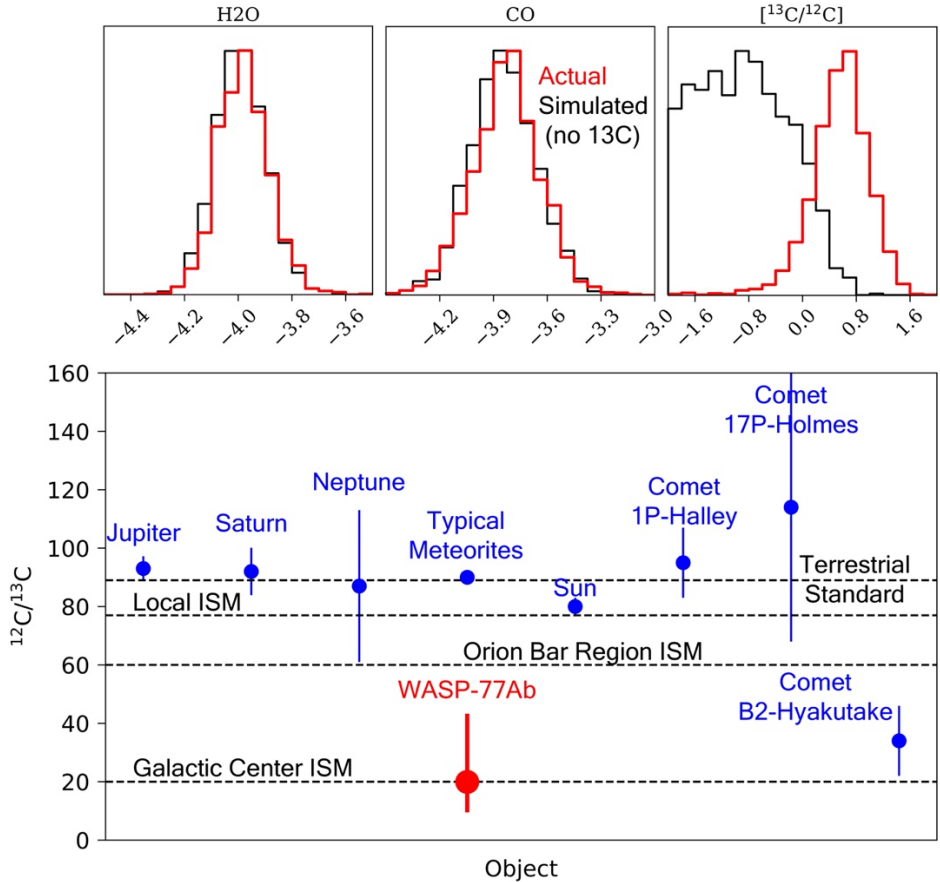

**ED Fig. 7**: Carbon isotopic abundance analysis. The top row of histograms compares the constraints from a nominal simplified retrieval model applied to the the true data (red) and an the reverse-injected data re-injected with $^{13}$C isotope removed model (black). The upper limit on the simulated data and bounded constraint arising from the true dataset suggests that there is indeed isotopic information in these IGRINS data. The bottom panel compares the retrieved $^{12}$C to $^{13}$C ratio (red) to common solar system bodies (blue, after Ref.[73]) and various reference values (galactic interstellar medium (ISM) components, and Earth (terrestrial), black dashed lines). WASP-77Ab sits anomalously low (enhanced $^{13}$C) compared to most solar system objects.